\documentclass{aa}
\usepackage{natbib}
\usepackage{graphicx}
\begin{document}

\title{Radio Variability of the  Sagittarius 
$ {\rm A^*}$ due to an Orbiting Star}

\author{Heon-Young Chang\inst{1}
          \and
        Chul-Sung Choi
          \inst{2}
}

\offprints{Heon-Young Chang}

\institute{
Korea Institute for Advanced Study \\
207-43 Cheongryangri-dong Dongdaemun-gu, Seoul 130-012, Korea\\
\email{hyc@ns.kias.re.kr}
\and
Korea Astronomy Observatory \\
36-1 Hwaam-dong, Yusong-gu, Taejon 305-348, Korea\\
\email{cschoi@kao.re.kr}
}

\date{Received ; accepted }

\abstract{
Recently, unprecedentedly accurate data on the orbital motion of stars 
in the vicinity of  the  Sagittarius  $ {\rm A^*}$ have been available.
Such information can be used not only to constrain the mass of the supermassive
black hole (SMBH) in the Galactic center but also to study the source 
of the radio emission. Two major competing explanations
of the radio spectrum of  the  Sagittarius $ {\rm A^*}$ are based on 
two different models, that is, hot accretion disk and jet.
Hence, independent observational constraints
are required to resolve  related issues.
It has been suggested that a star passing-by a hot accretion disk
may cool the hot accretion disk by  Comptonization 
and consequently cause the radio flux variation.
We explore the possibility of using the observational data of the star 
S2, currently closest to the Galactic center, to distinguish
physical models for the radio emission of the Sagittarius $ {\rm A^*}$, by
applying the  stellar cooling model to the  Sagittarius $ {\rm A^*}$ 
with the orbital parameters derived from  the observation. The
relative difference in the electron temperature due to the stellar cooling from
the star S2 at the pericenter is a few parts of a thousand
and the consequent relative radio luminosity 
difference is order of $10^{-4}$. Therefore, one could possibly expect to observe
the radio flux variation with a periodic or quasi-periodic 
modulation  in the frequency range  at $\nu \la {\rm 100 MHz}$
if the radiatively inefficient hot accretion flows 
are indeed responsible for the 
radio emission, contrary to the case of a jet.
According to our findings,  even though no periodic radio flux variations
 have been reported up to date a radiatively inefficient hot 
accretion disk model cannot be conclusively ruled out. This is simply because
the current available sensitivity is insufficient and because
the energy bands they have studied are too high 
to observe the effect of the star S2 even if
it indeed interacts with the hot disk.
}

\authorrunning{Chang \& Choi}
\titlerunning{Radio Variability of the  Sagittarius 
$ {\rm A^*}$ due to a Star}
\maketitle
  
\section{Introduction}

The compact radio source in our Galactic center
Sagittarius $ {\rm A^*}$ is widely believed to be associated
with an accreting supermassive black hole (SMBH) 
whose mass is $\sim 10^{6}M_\odot$ (Eckart \& Genzel 1997;
Ghez et al. 1998; Melia \& Falcke 2001; Eckart 2002). 
A number of models for the observed radio spectrum are essentially
based on an accretion process as quasars and active
galactic nuclei are powered by accreting SMBHs 
(e.g., Rees 1984). Though the existence of the SMBH at the Galactic center and 
its role seem unanimously accepted, the details of the 
accretion process and/or the nature of the central inner part of the accretion
flow remain unsettled. For instance, even the
recent {\it Chandra} observation of X-ray flare
by Baganoff et al. (2001) have been explained by
physically quite different  models of the Sagittarius  $ {\rm A^*}$ 
(Markoff et al. 2001; Liu \& Melia 2002; Yuan, Markoff, \& Falcke 2002).

Lower radio luminosities from Sagittarius  $ {\rm A^*}$
can be reasonably well explained by  the radiatively
inefficient accretion flow, such as, 
 advection-dominated accretion flows (Narayan, Yi, \& Mahadevan 1995).
The radiative luminosity of the advection-dominated accretion flows 
(ADAFs) is much less than 
that of the standard thin disk (Shakura \& Sunyaev 1973).  
The ADAFs have a low luminosity since 
most of the energy in the flows is stored in hot ions 
and advected
into the central black hole due to the low efficiency 
of heat transfer from ions to 
electrons (Ichimaru 1977; 
Rees et al. 1982; Narayan \& Yi 1994). 
The electron temperature is very high, 
and thus the electrons are  relativistic. 

Several models have been further introduced to account 
for the detailed spectrum of the Sagittarius $ {\rm A^*}$. Other versions of the
radiatively inefficient accretion flow, for instance, 
are accretion flows with significant macroscopic convection
(Narayan, Igumenshchev \& Abramowicz 2000; Quataert \& Gruzinov 2000a),
so called convection-dominated accretion flows (CDAF), those with mass loss
due to outflows from the accretion 
flow (Blandford \& Begelman 1999; Turolla \& Dullemond 2000), 
advection-dominated inflow-outflow solutions (ADIOS). A truncated disk with
a radio jet has been also 
proposed (Falcke \& Biermann 1999; Falcke \& Markoff 2000; 
Beckert \& Falcke 2002; Yuan, Markoff, \& Falcke 2002). 
They have both virtues and drawbacks in explaining the spectrum in details,
implying that the spectral energy distribution alone is probably insufficient
to sort out models.
Hence, the independent observations
resolving the central part are required to settle down related issues.
One of examples is the polarization observation.  Constraints by
the linear/circular polarization measurements seem quite robust
(Agol 2000; Quataert \& Gruzinov 2000b; Melia, Liu, \& Coker 2000;
Aitken et al. 2000; Bower et al. 2003). 
The measured polarizations provide information
on the emitting region, the mass accretion rate, the nature of the accretion
flows, the physical process of the radio emission, and so on.
Another example which has been  of great interests
for observers is flux variations (Baganoff et al. 2001; Duschl \& Lesch 1994; 
Eckart 2002; Hornstein et al. 2002; 
Goldwurm et al. 2003; Zhao, Bower, \& Goss 2001; Zhao et al. 2003).  
The short time-scale variation is of course important 
since it may provide the information on the physical 
change in the inner part of the accretion 
disk (e.g., Mushotzky, Done, \& Pounds 1993). 
On the other hand, the long time-scale variation in the very low frequency
band can also provide the useful
information on the central engine of the source like the Sagittarius $ {\rm A^*}$.

This paper is motivated by the recent observational report 
on the proper motion of stars close to 
the  Sagittarius $ {\rm A^*}$  (Sch\"odel et al. 2002; 
Gezari et al. 2002; Ghez et al. 2003).
Stellar proper motion data covered by an interval from 1992 to 2002 
allows to determine orbital accelerations for some of the most central 
stars of the Galaxy, thus the mass of the  Sagittarius $ {\rm A^*}$. 
The observations covering
both pericenter and apocenter passages show that the star S2, 
currently closest star to the  Sagittarius  $ {\rm A^*}$,  is on a bound, 
highly elliptical Keplerian orbit around the SMBH, 
with an orbital period of $\sim 15$ years, 
a pericenter distance of only 124 AU, or $\sim 2000$ Schwarzschild radius, 
the eccentricity of 0.87. As mentioned above, 
it will be interesting to consider a way to constrain the central source 
of the Sagittarius  $ {\rm A^*}$ using  this kind of observation.
We explore the possibility of using the observational data to distinguish
physical models for the radio emission of the Sagittarius $ {\rm A^*}$.
A similar attempts to this paper have been already
made in the sense that the same observational data 
are used to test putative accretion disk theories 
(Nayakshin \& Sunyaev 2003; Cuadra, Nayakshin, \& Sunyaev 2003).
Using the three dimensional 
orbit of the star S2
the latter authors 
concluded that there could exist no optically thick 
and geometrically thin disk near
the Sagittarius  $ {\rm A^*}$ or the cool disk must have
a large inner edge
since otherwise it  should have exhibited observational
signatures over the period of the observation campaign. 
In this paper we  
apply the stellar cooling model suggested by Chang (2001) 
to the particular case of the  Sagittarius $ {\rm A^*}$
by recalculating with the orbital parameters derived from 
the observation. 

\section{Stellar Interaction with ADAF}

A hot accretion disk is believed to exist in low luminosity AGNs
and dormant galaxies, such as, our own 
Galaxy (Narayan, Yi, \& Mahadevan 1995; Narayan et al. 1998; Ho 1999). 
If there is an accretion disk around the SMBH, several processes
may occur due to the interaction of a flying-by star and the accretion
disk around the SMBH (Syer, Clarke, \& Rees 1991; 
Hall, Clarke, \& Pringle 1996), without mentioning the tidal disruption
events (Cannizzo, Lee, \& Goodman 1990; Rees 1988, 1990; 
Menou \& Quataert 2001; Komossa 2002; Choi et al. 2002). 
A  more interesting
phenomenon can be observed  particularly 
when the accretion disk is relativistically hot. 
Stellar interactions with such a hot accretion flow around 
the SMBH play an intriguing role in that an flying-by star 
may cool the hot accretion disk as a result of Comptonization (Chang 2001). 
One of observable signatures of a flying-by star in 
the hot accretion disk, e.g., the ADAFs is
the decrease of the electron temperature and subsequently 
the radio flux of the hot accretion disk. 
In the followings it is briefly summarized what happens when a bright star
encounters a hot accretion disk. 

Firstly, when 
a star passes through the accretion disk around the SMBH the
dynamical friction causes the viscous heating. The power is given by 
$P = F_{\rm df}~  v_{\rm rel}$, where $F_{\rm df}$ is the drag 
force and $ v_{\rm rel}$
is the relative velocity of the star with respect to the background gas.
The drag force $F_{\rm df}$ on a star with mass $M_*$
moving through a uniform gas density $\rho$ with 
the relative velocity $v_{\rm rel}$
can be estimated  as 
\begin{equation}
F_{\rm df}=-4 \pi I \biggl(\frac{GM_*}{v_{\rm rel}} \biggr)^2  \rho,
\end{equation}
where the negative sign indicates that the force acts in the opposite 
direction of the star, $G$ is the gravitational 
constant (Ostriker 1999; Narayan 2000). 
The  coefficient $I$ depends on the Mach number, 
${\cal M} \equiv v_{\rm rel}/c_{\rm s}$, 
where $c_{\rm s}$ is the sound speed of the medium. 
In the limit of a slow moving ${\cal M} \ll 1$, $I_{\rm subsonic} \rightarrow 
{\cal M}^3/3$, so that the resulting $F_{\rm df}$  
is proportional to the relative
speed of the star. In the limit of a fast moving ${\cal M} \gg 1$, 
$I_{\rm supersonic} \rightarrow \ln(v_{\rm rel}t/r_{\rm min})$, 
where $r_{\rm min}$ is 
the effective size of the regime where the gravity of the star dominates. 
We take the supersonic 
estimate of $I$, as it gives an upper limit on the heating due to the drag force,
which makes the total cooling estimate obtained here a lower limit of
the stellar cooling effect.

Secondly,
on the other hand, the stellar emission may cool the gaseous medium. 
In the ADAFs a star and its
motion may enhance the cooling by bremsstrahlung and Comptonization 
processes.
The gas density in front of the  star may  be increased 
as the motion of the star may compress the gas. The
bremsstrahlung cooling rate per volume is increased as the density 
increases  proportionally to the square of the density (Stepney \& Guilbert 1983). 
Comptonization is also possible because the electrons in the ADAFs are 
relativistic since radiation emitted by the star is an important 
source of soft photons.
The stellar cooling rate due to Comptonization becomes 
relatively important than those due to other processes of accretion disk cooling 
when the mass accretion rate becomes small.

\begin{figure}
\resizebox{\hsize}{!}{\includegraphics{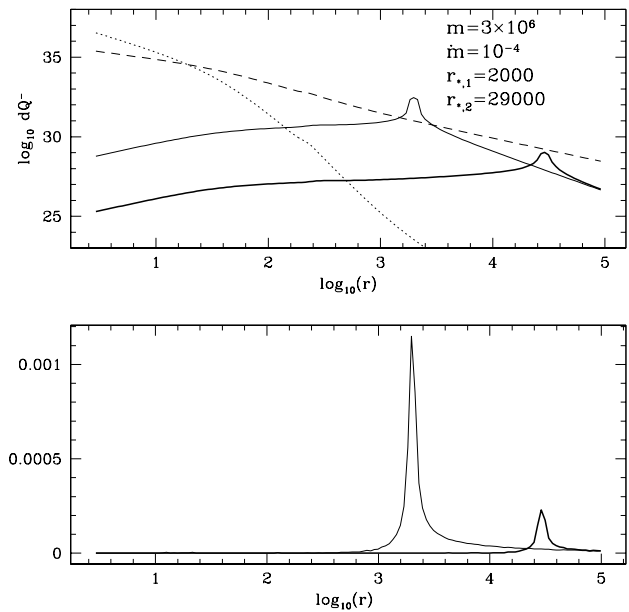}}
\caption{Upper panel: 
The volume integrated cooling rate over the spherical shell 
due to various cooling mechanisms are shown as a function of $r$ in log scales.
The $dQ^{-}$'s are in $ {\rm ergs~s^{-1}}$. 
Lower panel:
The relative electron temperature differences are shown
as a function of $r$. 
The relative electron temperature difference is defined as $(T_0-T_*)/T_0$,
where $T_0$ is the electron temperature of 
the case without the stellar cooling.
The thin and thick solid curves represent
the cases when the cooling star is at $r \sim 2000$
and  $r \sim 29000 $, respectively. 
The dotted curve and the dashed curve represent volume-integrated cooling rate 
due to synchrotron cooling and
 bremsstrahlung cooling, respectively. 
}
\end{figure}

\section{Radio Variation due to Stellar Cooling}

Making the simplest assumption of an optically thin and quasi-spherical 
hot accretion disk, 
we examine observational features  due to the hot accretion flow
present around the  Sagittarius $ {\rm A^*}$
through its interaction with the closely flying-by star, that is, S2. 
Using the orbital parameters and the observed positions of 
the currently closest star S2 (Sch\"odel et al. 2002; Ghez et al. 2003), 
we are able to calculate the radio flux variation 
which could have been observed with a similar period
of the orbital period of the star.
We adopt the following dimensionless variables
throughout the paper : mass of 
the SMBH $m=M/M_\odot$; radius from the SMBH $r=R/R_g$, 
where $R_g=2GM/c^2=2.95 \times 10^5~ m~{\rm cm}$; 
and mass accretion rate $\dot{m}=\dot{M}/\dot{M}_{\rm Edd}$,
where $\dot{M}_{\rm Edd}=L_{\rm Edd}/\eta_{\rm eff} c^2= 
1.39 \times 10^{18}~ m~ {\rm g~ s^{-1}}$ (the Eddington 
accretion rate assuming $\eta_{\rm eff}=0.1)$. We model the accreting
gas as a two-temperature plasma.
As typical values in a model 
for the ADAFs parameters are taken to be $r_{\rm min}=3$, $r_{\rm max}=10^5$,
$\alpha=0.3$,  $\beta=0.5$, and $\delta=0.0$ (see, e.g., Narayan \& Yi 1995; 
Quataert \& Narayan 1999).

\begin{figure}
\resizebox{\hsize}{!}{\includegraphics{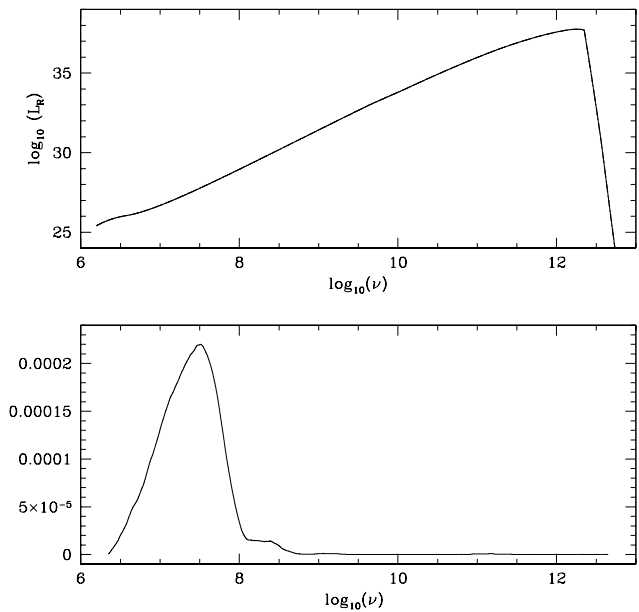}}
\caption{Upper panel: 
The radio spectrum of the ADAFs is shown.  
We show the radio spectra of the ADAFs without the stellar cooling by
the solid curve, and with the stellar cooling at the pericenter
by the dotted curve and at the apocenter by the dashed curve.
Note that all the curves are indistinguishable at this scale.
The luminosity  in $ {\rm ergs~s^{-1}}$ and 
the frequency in Hz are shown in the log scale.
Lower panel: The relative difference of 
the radio spectrum at two different epochs is shown.
}
\end{figure}

Provided that the background gas environment is described by the ADAF model,
the volume-integrated cooling rate due to stellar emission
$dQ^{-}_{\rm *,C}$ over the spherical shell at $r$ can be obtained.
We plot $dQ^{-}_{\rm *,C}$ with other volume-integrated cooling rates 
as a function of $r$ in Figure 1.
We adopt the bolometric luminosity of S2 as  that of an O8 dwarf
(Ghez et al. 2003).
The dotted curve and the dashed curve represent volume-integrated cooling rate 
due to synchrotron cooling $dQ^{-}_{\rm sync}+dQ^{-}_{\rm sync,C}$ and
 bremsstrahlung cooling $dQ^{-}_{\rm br}+dQ^{-}_{\rm br,C}$. 
The thin and thick continuous curve
represent $dQ^{-}_{\rm *,C}$ when the S2 star is at pericenter and at apocenter,
respectively. 
These volume-integrated cooling rates are subject to the mass 
accretion rate to the central SMBH. The  synchrotron cooling $dQ^{-}_{\rm sync}$ 
and bremsstrahlung cooling $dQ^{-}_{\rm br}$ are reduced as the mass accretion
rate is decreased. The stellar cooling rate behaves similarly.
However, its relative contribution becomes more significant compared with
others as the mass accretion rate is small.
We adopt the mass accretion rate is $\dot{m} = 10^{-4}$ (Quataert, Narayan, \& 
Reid 1999; Quataert \& Gruzinov 2000b), which
corresponds to the  favored accretion rate estimation from the observation
when the  ADAF model is assumed.
We also show the relative differences in the electron temperature
as a function of $r$ when the cooling star is at $r \sim 2000$
and  $r \sim 29000$ denoted by the thin and thick solid curves, respectively. 
The relative electron temperature difference  is defined as $(T_0-T_*)/T_0$,
where $T_0$ is the electron temperature of the case without the stellar cooling.
The electron temperature is again averaged over the volume of the shell.
As shown in the plot, for a given SMBH mass and the mass accretion rate
the suppression of the temperature due to the stellar cooling becomes
less significant as the cooling star is at farther away from the central SMBH.

In Figure 2, we show the radio spectrum of the ADAFs in the upper panel
and the relative difference of the radio spectrum due to
the stellar coolings at two different epochs.
The relative difference of radio spectrum has been given for two
different epochs, that is, when the star is at pericenter and at
apocenter.
Since the dominant effects on the spectrum is due to the inner parts of 
the ADAFs, the stellar cooling at farther from the SMBH changes the spectrum 
less significantly.  The suppression of the radio
spectrum due to the stellar cooling is the greatest at the frequency corresponding
to the position where the star cools the accretion disk  (see Mahadevan 1997). 
It can be
understood by the fact that the synchrotron radio emission of the ADAFs at 
 each frequency is closely related to a specific radius. For instance, the 
emission at higher frequencies originates at smaller radii, 
or closer to the central supermassive black hole.
As shown in the lower panel Comptonization
of stellar soft photons from the star  at  $r \ga 10^3$ affects the 
radio spectrum at $\nu \la {\rm 100 MHz}$.

\section{Summary and Discussions}

When a star like S2 interacts with the hot accretion disk such as the ADAFs
in the Galactic center
one would expect many interesting effects. One of observable
signature of a stellar encounter with 
the hot accretion disk such as the ADAFs is 
the depression of the radio flux due to the stellar cooling, whose 
variation could show periodic or quasi-periodic modulation.
We have attempted to calculate what one may actually expect using
observed parameters of the currently closest star S2. The
relative electron temperature difference is a few parts of a thousand
without and with the stellar cooling in the case
when the star S2 is near at the pericenter.
Subsequently the  radio spectrum  shows the 
suppression of the radio
spectrum due to stellar cooling  which is the greatest at the frequency
at $\nu \la {\rm 100 MHz}$ for  stellar soft photons 
from the star at $r \ga 10^3$.
The relative radio luminosity difference without and with the stellar
cooling is small, order of $10^{-4}$.
Bower et al. (2002) have reported  
multiepoch, multifrequency observations of the Sagittarius  $ {\rm A^*}$, 
from 1981 to 1998, of which  data
have been taken at  1.4, 4.8, 8.4, and 15 GHz bands. They have found
no periodic radio flux variation with a period $\sim $ 15  years,
which is naturally expected from the presence of a hot disk.
We suggest that this observation cannot be used yet to distinguish two
competing models, i.e., hot accretion disk and jets. 
That is, even though no periodic radio flux variations
 have been found in the observations  a radiatively inefficient hot 
accretion disk model cannot be conclusively ruled out. This is simply because
the currently available sensitivity is insufficient and because
the energy bands they have studied are too high 
to observe the effect of the star S2 even if
it indeed interacts with the hot disk.

We tentatively 
conclude that even the currently closest pass of the star S2 is insufficiently
close enough to meaningfully constrain the nature of the 
Sagittarius  $ {\rm A^*}$. Yet again, we would like to emphasize
that currently available data are out of range which the star S2 would
have affected. 
Quantitative implications may be subject to 
parameters we adopt for the background accretion flow model and
the physical parameters of the encountering star.
One may employ another version of the
radiatively inefficient accretion flow, for instance, CDAFs.
Changing the background from ADAFs to CDAFs has insignificant modification
in our conclusions since the convection in the hot accretion flows alters
the very inner part of the accretion flows and therefore the spectrum
at high frequency range,  
$\nu \ga {\rm 1 GHz}$ (see Ball, Narayan, \& Quataert 2001). The mass
loss in ADIOS also causes a significant effect only at higher frequency range
than we have interests. We, however, point out that a long monitoring
of radio flux and the stellar proper motion observation are still
worthwhile to be pursued since a hypothetical star orbiting a very
eccentric orbit might exist near at its apocenter where
it spends most of the orbital period and yet more closely pass by the 
Sagittarius  $ {\rm A^*}$ than the star S2.
If that happens we may observe the radio flux variation at 
$\nu \ga {\rm 1 GHz}$, where one should be more careful in choosing
the background accretion flow model. One may also attempt to monitor
LLAGNs, which are believed to host the ADAFs or their variations (Ho 1999).
For an accreting SMBH
with a lower mass accretion rate and more closely flying-by star may 
exhibit their existence.

\acknowledgements{
We would like to thank the anonymous referee for positive
comments.}

\end{document}